\documentclass[12pt]{article}

\def\refitem#1{\relax}
\def\refitem#1{\relax}

\newcommand{\mycite}[1]{~{\cite{#1}}}

\newcommand{\ac}{\mathcal{A}} \newcommand{\Kr}{\mathcal{W}}
\def\alps{\relax\ifmmode\alpha_s\else{$\alpha_s${ }}\fi}
\usepackage{amsmath} \usepackage{amssymb}

\newcommand{\beq}{\begin{equation}}
\newcommand{\eeq}{\end{equation}}                                  
\newcommand{\bd}{\begin{displaymath}}\newcommand{\ed}{\end{displaymath}}
\newcommand{\bit}{\begin{itemize}}\newcommand{\eit}{\end{itemize}} 
\newcommand{\ben}{\begin{enumerate}}\newcommand{\een}{\end{enumerate}}
\newcommand{\baa}{\begin{array}{lll}}\newcommand{\eaa}{\end{array}}
\newcommand{\ba}{\begin{eqnarray}}\newcommand{\ea}{\end{eqnarray}} 
\newcommand{\bmp}[1]{\begin{minipage}{#1}}\newcommand{\emp}{\end{minipage}}%
\def\eV{\relax\ifmmode{\rm e\kern-0.12em V}\else{\rm e\kern-0.12em V{ }}\fi}
\def\MeV{\relax\ifmmode{\rm M\eV}\else{\rm M\eV{ }}\fi}
\def\GeV{\relax\ifmmode{\rm G}\eV\else{\rm G\eV{ }}\fi}            
\def\MSbar{\relax\ifmmode\bar{\rm MS}\else{$\bar{\rm MS}${ }}\fi}  
\def\as{\relax\ifmmode \alpha_s\else{$ \alpha_s${ }}\fi}           
\def\al{\relax\ifmmode\alpha\else{$\alpha${ }}\fi}
\def\alps{\relax\ifmmode\alpha_s\else{$\alpha_s${ }}\fi}
\def\msbar{\relax\ifmmode\overline{\rm MS}\else{$\overline{\rm MS}${ }}\fi}
\def\albar{\relax\ifmmode{\bar{\alpha}}\else{$\bar{\alpha}${ }}\fi}
\def\albarE{\relax\ifmmode{\bar{\alpha}_E}\else{$\bar{\alpha}_E${ }}\fi}
\def\alphaE{\relax\ifmmode{\alpha_E}\else{$\alpha_E${ }}\fi}
\def\albarEQ{\relax\ifmmode{\albar_E(Q^2)}\else{$\albar_E(Q^2)${ }}\fi}
\def\albarM{\relax\ifmmode{\bar{\alpha}_M}\else{$\bar{\alpha}_M${ }}\fi}
\def\albars{\relax\ifmmode{\bar{\alpha}_s}\else{$\bar{\alpha}_s${ }}\fi}
\def\albarsQ{\relax\ifmmode{\bar{\alpha}_s(Q^2)}\else{$\,\bar{\alpha}_s(Q^2)${}}\fi}
\def\agoth{\relax\ifmmode{\mathfrak A}\else{$\,{\mathfrak A}${ }}\fi}
 \def\agothk{\relax\ifmmode{\mathfrak A}_k\else{${\mathfrak A}_k${ }}\fi}
\def\agothks{\relax\ifmmode{\mathfrak A}_k(s)\else{${\mathfrak A}_k(s)${}}\fi}
\def\acal{\relax\ifmmode{\cal A}\else{${\cal A}${ }}\fi}
  \def\acalk{\relax\ifmmode{\cal A}_k\else{${\cal A}_k${ }}\fi}
 \def\acalkQ{\relax\ifmmode{\cal A}_k(Q^2)\else{${\cal A}_k(Q^2)${ }}\fi}
\def\alphaMs{\relax\ifmmode\alpha_M(s)\else{$\alpha_M(s)${ }}\fi}
\def\alphaEQ{\relax\ifmmode{\alpha_E(Q^2)}\else{$\alpha_E(Q^2)${ }}\fi}
\def\alphaM{\relax\ifmmode{\alpha}_M\else{$\alpha_M${ }}\fi}
 \newcommand{\Ups}{\Upsilon}
\def\Lms{\Lambda_{\msbar}}
\begin{document}

\begin{center}
 \large{\bf Analytic Perturbation Theory for QCD
 \protect\\ Practitioners and Upsilon Decay}\\
 \bigskip
 D.~V.~Shirkov$^\dagger$ and
 A.~V.~Zayakin$^\ddagger$\footnote{E-mail:
 zayakin@theor.jinr.ru } \\
 \medskip{\small \it
 $\dagger$ Joint Institute for Nuclear Research, Dubna, 141980,
 Russia\\
 $\ddagger$ Moscow State University, Moscow 119992,
Russia }\medskip
\end{center}

\begin{abstract}
  Within the ghost-free Analytic Perturbation Theory
 (APT), devised in the last decade for low energy QCD,
 simple approximations are proposed for 3-loop analytic
 couplings and their effective powers, in both the
 space-like (Euclidean) and time-like (Minkowskian) regions, accurate enough in
 the large range (1--100 GeV) of current physical interest.\par
  Effectiveness of the new Model is illustrated by the
 example of $\Ups(1\mathrm{S})$ decay where the standard
 analysis gives $\alps(M_{\Ups})=0.170\pm 0.004$ value
 that is inconsistent with the bulk of data for $\alpha_s$.
 Instead, we obtain $\as^{Mod}(M_{\Ups})=0.185\pm 0.005$
 that corresponds to $\as^{Mod}(M_Z)=0.120\pm 0.002\,$
 that is close to the world average.\par
  The issue of scale uncertainty for $\Ups$ decay is
 also discussed.
 \end{abstract}

 \section*{Introduction}
 Theoretical expressions for measured quantities of
 hadron physics contain QCD running coupling $\as\,.$
 The common formula (eq.(7) in Bethke\cite{Bethke:2000ai}
 or eq.(9.5) in Particle Data Group (PDG)
 review\cite{pdg04}) suffers from the unphysical pole
 singularity that is an indispensable feature of all
 renormalization group (RG)sums of ultra-violet(UV) logs
 obtained from perturbation theory. In QCD, this issue
 becomes especially troublesome in the few GeV region.\par

  The first model of a ghost-free QED coupling
 $\alpha_{an}^{QED}(Q^2;\alpha)\,$ was devised long
 ago~\cite{BLSh59} on the basis of the K\"allen--Lehmann
 analyticity in the complex $Q^2\,$ plane. The price for
 the absence of ghost is non-analyticity of
 $\alpha_{an}^{QED}(Q^2;\alpha)\,$ with respect to
 coupling constant $\alpha\,$ at $\alpha=0\,.$\par

  The idea to use the K\"allen--Lehmann imperative to
 get rid of unphysical singularities (like Landau pole)
 was transferred to QCD in the mid-90s\cite{ShSol96} and
 named ``analyticization". It results in a regular in
 the low-$Q^2\,$ Euclidean domain effective coupling
 function with a finite IR limit $\alphaE(Q^2=0)\,.$\par

  On this basis, another ghost-free construction for QCD
 effective coupling in the Min\-kowskian domain
 $\alpha_M(s)$ was introduced \cite{Milton:1996fc} via
 integral transformation. This result turned out to be
 equivalent to the result of $\pi^2$-terms summation
 derived by similar means\cite{Rad82,KrasPiv82} in the
 early 80s (see also, \cite{Bjork89}).\par

  Later on, both the constructions were joined
 \cite{Sh01tmp} by suitable integral transformations
 into the so-called {\it Analytic Perturbation Theory}
 (APT). An essential feature of the APT scheme is that
 Minkowskian and Euclidean counterparts for powers of
 usual QCD coupling $(\alps(Q^2))^k\,$ form nonpower
 sets $\{\agothk(s)\}\,$ and $\{\acalk(Q^2)\}\,.$ For
 the fresh reviews of APT see~\cite{Sh01epjc,Sh05bari}.\par

  Meanwhile, ghost free APT expressions for effective
 couplings in the Euclidean \alphaEQ and Minkowskian
 \alphaMs regions, as well as for their ``effective
 powers" $\ac_k(Q^2)$ and $\agothk(s)$, are presented
 by simple analytic expressions only in the one-loop
 case (see eqs.(\ref{AE11})-(\ref{4m})), which are not
 accurate enough for practical goals.\par

   Higher-loop APT expressions are more intricate involving a
 special Lambert function. A few years ago the first three of them
 $\ac_k(Q^2)\,,$ $\agoth_k(s);\, \,(k=1,2,3)\,,$ sufficient for
 most of applications, were tabulated by Magradze and
 Kourashev~\cite{KourMagr01} in the 3-loop case. \par

  Here, we propose simple model expressions for 3-loop
 APT couplings $\alpha_E(Q^2)=\acal_1,\,\alpha_M(s)=
 \agoth_1\,$ and for higher expansion functions
 $\acalk(Q^2)\,,$ $\agoth_k(s)\,$ connected by simple
 iterative relations. The Model depends on one additional
 parameter and is valid in the region above 1 \GeV.\par

 The paper is organized as follows. Sections 1 and 2
 contain a brief review of the main ideas, technique and
 some results of APT. In Section 3, we present our Model.
 Section 4 contains revised analysis of $\Ups$ decay data
 with $\as$ value extracted anew by means of APT and our
 Model. Special attention is paid to scale uncertainty. The
 last, Section 5, is devoted to the summary of the results.

\section{\large Outline of the Analytic Perturbation
  Theory}  
  To start this short overview, remind that the
 cornerstones of APT are {\it the $Q^2$ analyticity of
 coupling functions} and {\it compatibility with linear
 integral transformations}.\par
 Here follows compendium of main definitions. The most
 elegant APT formulation is based on a set of spectral
 functions $\{\rho_i(\sigma)\}\,$ defined as
 \beq
 \rho_k(z)=\mathrm{Im}([\as(-z)]^k)\,\label{density}\eeq
 The first of them, $\rho_1=\rho(\sigma)\,$ is just
 K\"allen--Lehmann spectral density for the Euclidean
 APT coupling. Then, higher Euclidean (``analyticized
 $k$th power of coupling in the Euclidean domain'') and
 Minkowskian (``effective $k$th power of coupling in
 the Minkowskian domain") APT functions will be
 respectively given by
 \beq\label{min}
 \ac_k(Q^2)=\mathbb{A}[\as^k]=\frac{1}{\pi}
 \int^{+\infty}_0 \frac{\rho_k(\sigma)\,d\sigma}
 {\sigma+Q^2}\,;\quad\agoth_k(s)=\mathbb{R}[\as]
 =\frac{1}{\pi}\int_s^{+\infty}
 \frac{d\sigma}{\sigma}\rho_k(\sigma)\,\,.\eeq
 These functions are related by integral transformation
 $$
 \acal_k(Q^2)=\mathbb{D}[\agoth_k]=Q^2\int^{+\infty}_0
\frac{\agoth_k(s)\, d s}{(s+Q^2)^2}$$
 and its reverse. The differential relations connecting
 higher spectral functions
 \beq
 -\frac{1}{k}\frac{d\rho_k}{d\,\ln \sigma}=
 \beta_0 \,\rho_{k+1}+ \beta_1\,\rho_{k+2}+
 \,\dots\,,\quad\beta_0(n_f)=\frac{33-2 n_f}{12\pi}
 \,, \quad \beta_1=\frac{153-19 n_f}{24\pi^2}
 \,,\, \dots \label{beta0}\eeq
 induce analogous relations for expansion functions
 \begin{equation}\label{r-r}
 \frac{1}{k}\frac{d\agoth_k(s)}{d\,\ln s}=-
 \sum_{n\geq 1} \beta_{n-1}\agoth_{k+n}(s)\,,
 \qquad \frac{1}{k}\frac{d\acal_k(Q^2)}{d\,
 \ln Q^2}=-\sum_{n\geq 1}\beta_{n-1}\ac_{k+n}(Q^2),
 \end{equation}
 which can be used for iterative definitions.
 Numerically, these beta-coefficients (defined
 according to Bethke\cite{Bethke:2000ai}) and
 their useful combination $b=\beta_1/\beta_0^2\,$
 are of the order of unity\footnote{There is a misprint
 in numerical values of $\beta_0(4\mp 1)\,$
 in paper \cite{Sh01epjc}.}
\begin{equation}\label{beta}
 \beta_0(4\mp 1)=0.6631\pm0.0530\,,\,\, \beta_1(4\mp 1)
 =0.3251\pm0.0802\,,\,\, b(4\mp 1)=
 0.7392^{+0.0509}_{-0.0814}\,.\eeq

 \section{\large Main Results of APT }
 \underline{\sf One-loop case.} In this case, the
 APT formulae are simple and elegant. Starting
 with the perturbative RG-improved QCD coupling
 $\,\as^{(1)}(Q^2)= 1/(\beta_0\,l)\,,$ with
 the help of~(\ref{density}),(\ref{min}) one
 arrives at the ghost-free effective Euclidean
 \footnote{Note, we change the notation for arguments
 of the APT functions: $Q^2\to l\,$ and $s\to L\,.$}
 \beq\label{AE11}
 \ac^{(1)}_1(l)=\frac{1}{\beta_0}\left(\frac{1}{l}
 -\frac{1}{e^l-1}\right)=\frac{1}{\beta_0\pi}\left(
 \frac{1}{l}+\frac{\Lambda^2}{\Lambda^2-Q^2}\right)
 \,,\quad l=\ln\left(\frac{Q^2}{\Lambda^2}\right)\eeq
 and  Minkowskian
 \beq\label{AM11}  
 \agoth^{(1)}_1(L)=\frac{1}{\beta_0\pi}
 \arccos\left(\frac{L} {\sqrt{L^2+\pi^2}}\right)\,,
 \quad L=\ln\left(\frac{s}{\Lambda^2}\right)\eeq
 APT couplings. Here, higher functions $\ac_i\,
 ,\agoth_i\,$ can be defined via recursive
 relations (\ref{r-r}) with only one term in the
 r.h.s. The second and third one-loop functions are
 {\small
 \beq\label{2-3}  
 \begin{array}{c}\displaystyle
 \ac_2^{(1)}(l)=\frac{1}{\beta_0^2}\left(\frac{1}
 {l^2}-  \frac{e^l}{(e^l-1)^2}\right)\,,\quad
 \agoth_2^{(1)}(L)= \frac{1}{\beta_{0}^2}\,
 \frac{1}{L^2+\pi^2}\,;,\vspace{2mm}\\ \displaystyle
 \vspace{2mm} \ac_3^{(1)}(l)=\frac{1}{\beta_0^3}
 \left(\frac{1}{l^3} - \frac{1}{2}\frac{e^l+
 e^{2l}}{(e^l- 1)^3}\right) \,,\quad \agoth_3^{(1)}(L)
 =\frac{1}{\beta_{0}^3} \frac{L}{(L^2+\pi^2)^2}\,.
       \end{array}   \eeq }
 In Section 4.2, we shall also need the fourth
 Minkowskian  function,
 \beq\label{4m} 
 \agoth_4^{(1)}(L)= \frac{1}{\beta_{0}^4}\frac{L^2-
 \pi^2/3} {(L^2+\pi^2)^3}\,.\eeq

  All APT functions obey important properties that are
  valid in the higher-loop case:
 \begin{itemize}
 \item Unphysical singularities are absent with no
          additional parameters introduced.
 \item In the Euclidean and Minkowskian domains,
  QCD couplings $\alpha_E(Q^2)=\acal_1(Q^2)\,,$
 $\alpha_M(s)=\agoth_1\,$ and their ``effective powers"
   $\acal_k(Q^2)\,$ $\agoth_k(s)\,,$
 are different functions related by integral
 operations $\mathbb{A[\,\,]}$ and $\mathbb{R[\,\,]}\,$
 explicitly defined in eqs. (\ref{min}).
 \item Higher functions, e.g., (\ref{2-3}),(9) are not equal
    to powers of the first ones (\ref{AE11}), (\ref{AM11}).
 \item Expansion of an observable in coupling powers
  $(\as(Q^2))^n\,$ in the Euclidean or in $(\as(s))^n\,$
  in the Minkowskian case is substituted by nonpower
  expansion in sets $\left\{\acalk\right\}\,,$ or
  $\left\{\agothk\right\}\,$
  respectively. The latter expansions exhibit a faster
  convergence, as compared to common power expansion.
 \end{itemize} %

   Two particular notes are in order:\par
   --- The APT functions $\acal_k(Q^2)\,,\,\,\agoth_k(s)\,$
 essentially differ from common expansion functions
 $(\as)^k\,$ in the low energy region, where they are
 regular with finite IR limit. In particular,
 $\alpha_E(0)=\alpha_M(0)=1/\beta_0\,$. This behavior
 provides high stability with respect to variation of the
 renormalisation scheme \cite{ShSolPL97}. \par
 --- In the UV limit, all APT functions tend to their usual
 counterparts $(\as)^k\,.$ The related small parameters for
 the measure of deviation are $\epsilon_E\,$ in Euclidean
 and $\epsilon_M\,$ in Minkowskian case
\begin{equation}\label{epsil}
 \epsilon_E=\frac{\Lambda^2}{Q^2}\,\ln\left( \frac{Q^2}
 {\Lambda^2} \right)\,,\quad\epsilon_M= \frac{\pi^2}
 {\ln^2\left(s/\Lambda^2\right)}\,.\eeq
 Remark here that influence of APT contributions to QCD
 coupling and to observables is quite different in the
 Euclidean and Minkowskian regions. In the first case, APT
 correction $\sim \epsilon_E\,$ with rise of momentum
 becomes less than $1\%$ already at $Q \simeq 10\,\GeV.$
 On the contrary, the influence of Minkowskian correction
 $\epsilon_M\,$ can be traced up to $100\,\GeV\,$ scale
 where the difference between two-loop (NLO) analytic and
 non-analytic coupling makes roughly 5\%. Even in the 3-loop
 (NNLO) standard Minkowskian case, the effect of $\pi^2$
 terms $\sim \as^4\pi^2\beta^2_0\,$ remains
 essential\footnote{For more details see Sections 3.2 and
 4 in~ \cite{Sh01epjc}.} up to $10\,\GeV\,.$\smallskip

  \underline{\sf Higher-loop case.} The two-loop expressions
 are more complicated. Here, exact QCD coupling \as can be
 expressed explicitly
 \begin{equation}\label{bdef}
 \as^{(2)}(Q^2)=-\frac{\beta_0}{\beta_1}\frac{1}
 {1+\Kr_{-1}(z)}\quad \mbox{with}\quad z=-\frac{1}{b}
 \exp\left(-\frac{L}{b}- 1\right) \quad \mbox{and}
 \quad b=\frac{\beta_1}{\beta_0^2}\,.\eeq
 in terms of the Lambert function $\,\Kr(z)\,$ defined as a
 solution\footnote{One should be careful with choosing a
 proper branch $\Kr_{-1}$, see, e.g.,\mycite{Shirkov:1998sb}.}
 of the transcendental equation $\Kr e^{\Kr}=z\,.$ This
 expression yields rather involved formulae for $\ac_k$
 and $\agoth_k$ in terms of $\Kr_{-1}\,$.\par

  In the three-loop case, one encounters more complications.
 Here, only for Pad\'e approximated beta-function, exact
 solution can be expressed explicitly\mycite{KourMagr01} in
 terms of the Lambert function and one meets the same
 situation as with exact two-loop solution\footnote{For the
 popular two-loop expression that represents (like eq.(9.5)
 in~\cite{pdg04}) expansion of the iterative solution
  \vspace{-1mm}
 $$\label{it-app}
 \as^{(2,iter)}=\frac{1}{\beta_0 (l+b\ln l)}\,\simeq\,
 \as^{(2, appr)}(Q^2)=\frac{1}{\beta_0} \left(
 \frac{1}{l}- b\,\frac{\ln l}{l^2}\right)\,,$$
 the corresponding Minkowskian counterpart
 $\agoth_1^{(2, appr)}$ is known\mycite{Rad82}.
  However, numerically, it gives a rather crude
 approximation in the low-energy region.}.\par
   The total picture becomes even more involved after taking
 into account the matching relation for adjusting kinematic
 regions with different values of the flavour number
 $n_f\,.$ The devised scheme \cite{Sh01tmp}, known as
 ``global APT", has been studied by Magradze and Kourashev
 at the two- and three-loop level. They calculated numerical
 tables for the first three functions $\agoth_k{1,2,3}\,$
 and $\acal_{1,2,3}\,$ at three values of
 $\Lambda^{n_f=3}= 350,400,450\,\MeV\,$ in the interval
 $1\,\GeV <\sqrt{s},Q < 100\,\GeV\,$
 ~\cite{KourMagr01,Magr00hz}, and
 $\agoth_{1,2}\,,\,\acal_{1,2}\,$ in the interval
 $0.1\,\GeV<\sqrt{s},Q\lesssim 3\,\GeV\,$~\cite{magr03}.
 Unhappily, their numerical tables, as well as complicated
 analytic formulae, are not comfortable enough for QCD
 practitioners. Some other approximations proposed recently
 \cite{mss02} are not also of wide use yet. \smallskip

  \underline{Physical implications of APT.}
  Re-examination of various processes on the basis of APT
 during the last decade has been performed in both
 Minkowskian and Euclidean regions. For a five-years old
 review, see paper \cite{Sh01epjc}. Here, we list some
 fresh results.\par

  In Ref.\cite{BP-KSS04}, it has been shown that in using
 the standard perturbation theory for description of the
 pion electromagnetic form factor, the size of the NLO
 corrections is quite sensitive to the adopted
 renormalization scheme and scale setting. Replacing the
 QCD coupling and its powers by their APT images, both
 dependences are diminished and the predictions for the
 pion form factor turn out to be quasi-independent of
 scheme and scale settings.\par
  Applying appropriate generalization to the fractional
 powers~\cite{BMS05} of APT coupling, the authors of~
 \cite{BKS05} showed that the dependence of APT predictions
 for the pion form factor on the factorization scale was
 also diminished.\par

 Inclusive $\tau$-decay on the APT base has been studied
 by Solovtsov with co-authors. In paper \cite{MSSY:00},
 the stability of results with respect to the
 renormalization scale change
 was demonstrated. The QCD parameter value,
 $\Lambda_{f=3}\simeq 400\,\MeV\,,$ close to the world
 average was obtained in Refs.\cite{MSS:01,MSS:05} after
 due account of non-perturbative values of light quarks
 masses and summation of threshold singularities.\par

  One more curious application of APT appeared recently from
 the mass spectrum analysis of ground and first excited
 quarkonium states by the Milano group. There, it was
 argued\cite{Milan02,Milan04} that APT Euclidean coupling
 $\alphaEQ\,$ at the interval $Q\sim 100-400\,\MeV\,$ should
 take values corresponding to
 $\Lambda_{f=3}\simeq 375\,\MeV\,.$

 \section{Simple Model for 3-Loop APT Functions}
  \subsection{``One-Loop-Like'' Model}    
 Our aim is to construct simple and accurate enough (for
 practical use) analytic approximations for two sets of
 functions $\agoth_k$ and $\ac_k\,,k=1,2,3\,$. To reduce
 number of fitting parameters, one should better provide
 the applicability of the recurrent relations.

  To this goal, we suggest that one-loop APT expressions,
 eqs.(\ref{AE11}),(\ref{AM11}),(\ref{2-3}), with modified
 logarithmic arguments are used
 \beq\label{model} 
   \ac^{mod}_k(l)=\ac^{(1)}_k(l_*)\,;\quad
 \agoth^{mod}_k(L)= \agoth^{(1)}_k(L_*)\,,\eeq
 $L_*$ and $l_*$ being some ``two-loop RG times".

  Model functions~(\ref{model}) are related by the
 ``one-loop-type" recursive relations
 \begin{equation}\label{rec-mod}         
 \ac^{mod}_{n+1}=-\frac{1}{n\,\beta_0}
 \frac{d\ac^{mod}_n}{dl_*}=-\frac{1}{n\,\beta_0}
 \frac{d\ac^{mod}_n}{dl}\cdot \frac{d l}{d l_*}
 \,,\qquad \agoth^{mod}_{n+1}= -\frac{1}{n\,
 \beta_0}\frac{d\agoth^{mod}_n}{dL_*} \,.\eeq

 A simple expression for $l_*\,$ can be borrowed from
 \cite{SS-99}, where a plain approximation for the two-loop
 effective log in the Euclidean region was used
  $$
 \alpha_E^{mod}(Q^2)=\frac{1}{\beta_0}\left\{
 \frac{1}{l_2}+ \frac{1}{1-\exp(l_2)}\right\}\,,
 \quad l_2= l+b\ln\sqrt{l^2+4\pi^2}\,,$$
 with $b$ defined in (\ref{bdef}).
  The structure of $l_2$ was inspired there by an idea of
 compensation of the first complex branch-cut of the Lambert
 function arising in the exact two-loop solution. This
 approximation was shown to combine reasonable accuracy in
 the low-energy range with the absence of singularities for
 $\alpha_E\,.$ We extend this approach to higher functions
 in both the Euclidean and Minkowskian domains. To this
 goal, we change square root in ``effective logs" $L_2(a)\,$
 and $l_2(a)\,:\,\sqrt{l^2+4\pi^2}\to\sqrt{l^2+a\pi^2}\,$
 with $a\,,$ an adjustable parameter.
 It comes out from thorough numerical analysis that optimal
 value of the new parameter is $a\approx 2\,,$ while {\it
 effective boundaries between the flavor regions have to be
 chosen on quark masses} $m_c=1.3\,\GeV$ and $m_b=4.3\,\GeV$
 just as in the $\overline{MS}\,$ scheme. \par
  That is, {\sf we formulate our Model as a set of
 equations (12) with (6) -- (8) and }
 \begin{equation} \label{model2}
   L_*=L_2(a=2)=L+b\ln\sqrt{L^2+2\pi^2}\,,\quad
   l_*=l_2(2)=l+b\ln\sqrt{l^2+2\pi^2}\,.\eeq
  Here, $L\,$ and $l\,$ are defined via common $\Lms\,$
 values, like in (\ref{AE11}),(\ref{AM11}), for each of the
 flavor region. This choice provides us with overall
 accuracy of a few per cent  -- see Table 1.\par
  Advantage of Model (\ref{AE11})--(\ref{2-3}),(\ref{model}),
 (\ref{model2}) is that it involves only one new parameter,
 $a=2$ with $\Lms\,$ and $n_f$ taking their usual values.
\vspace{-3mm}

 \subsection{Accuracy of the Model vs data errors}
   In Table 1 we give the maximal errors of our Model
 expressions (\ref{2-3}), (\ref{model}),(\ref{model2}) in
 each $n_f\,$ range obtained by numerical comparison with the
 above-mentioned Magradze tables in the interval of 3-loop
 $\Lms^{(n_f=3)} \sim 350 - 400\,\MeV\,.$

  As it follows from the Table, errors of Model for the first
 three APT functions are small, being of an order of 1-2 per
 cent for the first functions, of 3-5\% for the second and of
 6-10\% for the third ones in the region above $1.5\,\GeV\,,$
 i.e., in the $n_f=4,5\,$ ranges. However, its accuracy
 in the $n_f=3\,$ region (above 1\GeV) is at the level of
 5-10 per cent.\par

  Meanwhile, relative contributions of typical LO, NLO and
 NNLO terms in APT nonpower expansion for observables
 are usually something like 60-80\%, 30-10\%, and 10-1\%,
 respectively (see Table 2 in Ref.\cite{Sh01epjc}).
  Due to this, the Model accuracy for many cases is defined
 by that of the first model functions $\acal_1^{mod}\,,\,
 \agoth_1^{mod}\,$, provided that QCD contribution to an
 observable starts from one-loop contribution $\sim \as\,.$
   At the same time, for quarkonium decays one meets the
 case with the leading contribution $\sim \as^3\,.$ There,
 the Model error is defined by accuracy of the third
 Minkowskian function $\agoth_3^{mod}\,$-- see Table 2.

  In this Table we compare our Model errors with some data
 errors in the low energy region. The last ones are taken
 from a recent Bethke's reviews\mycite{Bethke:2000ai}.

  With due regard for data errors, we can now set some
 total margin of accuracy that our Model would satisfy.
 This margin may be chosen, e.g., as 1/3 of the
 experimental error bar, which is no less than 10\%
 -- see column 5 in Table 2). Then the accuracy limit,
 imposed upon the first APT functions could be about 3\%,
 for the second function will be at least 3 times
 weaker than for the first one (say, 10\%), and for the
 third ones, at least 6 times weaker (say, 20\%).
 Analysing Table 1 according to these requirements, one
 concludes that it is not reasonable to use the Model below
 0.5 \GeV, whereas above it (or, in part of 3-flavour and in
 the 4- and 5-flavour regions) it is fully advisable. Now
 we may proceed to its practical application.\par

\section{$\Upsilon$(1S) Decay Revisited}

 To this goal, we take the $\Upsilon(1S)$ non-radiative
 decay. The main reason of this choice is the troubling
 disagreement,exceeding three standard deviations, between
 the coupling \as extracted from this $\Ups$ decay and the
 world average. This can be clearly seen from Fig. 9.2 of
 the Particle Data Group review~\cite{pdg04}.\par

  An observable, which could be an apt illustration of the
 proposed Model should obey the following criteria: first,
 influence of the $\pi^2$ terms upon its value should be
 large enough, this observable should be measured up to a
 sufficient precision. The $\Ups(1S)$ non-radiative decay
 satisfies these both. Indeed, on the one hand, the parameter
 $\epsilon_M\,$ defined in (\ref{epsil}) is not
 small\footnote{The importance of the $\pi^2\,$ terms in
 $\Ups$ decay was demonstrated recently. The rough estimate
 obtained in Section 4 of paper\cite{Sh01epjc} gave
  $\triangle\as(M_Y)\simeq+0.012\,$
 that is more than 5 \% correction to the NLO result.}
 in the region (3-5 GeV) related to this decay: \
 $\epsilon_M^{(2)}=\pi^2/L^2_2 \simeq 0.3 -0.2\,.$ \par

    On the other hand, the non-radiative decay of $1S$-state
 provides the best data precision ($2.0\%$  for the ratio
 of hadronic and leptonic widths~\cite{pdg04}). Decays of
 2S and $3S$ states give a poorer accuracy of this ratio.
   Besides, there is one more argument to re-examine $\Ups$
 decay; the essential disagreement\footnote{This issue is
 absent in Bethke's reviews \cite{Bethke:2000ai}, because
 there an $\as$ estimate is based only upon paper
 \cite{Pivov98a}, where a reasonable $\as$ value was
 extracted from $\Ups$ sum rules with Coulomb resummation
 taken into account.} with the world average, just mentioned
 above. \par

  At the same time, quarkonium decays are not the best proving
 ground for the test of model expressions we have devised.
 Indeed, in the choice of the Model we made accent on the
 accuracy of first APT functions. Meanwhile, these decays are
 described by expressions $\sim \as^3\,.$ Nevertheless, even
 with 7 per cent error (as it is seen in Table 2) the Model
 application to $\Ups(1S)$ decay will be interesting enough.

 \subsection{$\Ups$  Widths}
  The paper~\cite{KrasPiv82} was the predecessor of present
 work in revisiting $\Ups$ decay. 25 years ago, these
 authors showed the effect of analytical continuation to be
 significant for $\Ups$ decay width, resulting in a
 considerably larger $\Lms\,$ (1.5 times) than the values
 obtained earlier by standard RG improved perturbation
 theory. Later on, the importance of $\pi^2$ terms was
 emphasized by Bjorken \cite{Bjork89}. Nevertheless, to our
 knowledge, extraction of \as value was performed so far
  without due regard for the analytic continuation
 effects, with some exceptions mentioned below.\par

  A few words on other processes involving $\Ups$, like its
 radiative decays and $\Ups$ production. All these have low
 data precision (typically, radiative widths are measured
 with the $10\%-40\%$ accuracy). For an extensive review of
  $q\bar{q}\,$ decay widths see \mycite{Bodwin95}.

  The NLO ratio of hadronic and leptonic decay widths of the
 $\Ups$ $S$ state was given in~\cite{Mackenzie81} (see
 also~\cite{kobelthesis}) \footnote{This formula was
 reproduced in review~\cite{pdg04} with error, but
 corrected in the last online PDG version.}
 {\small
 \beq\label{Mack}
 R_\Ups=\frac{\Gamma\left(\Ups\to hadrons\right)}
 {\Gamma\left(\Ups\rightarrow e^+ e^-\right)}=
 \frac{10(\pi^2-9) \alpha_s^3(\mu)}{9\,\pi\alpha^2
 (M_\Ups)} \left [1+\frac{\alpha_s(\mu)}{\pi}
 \left(\tilde{\beta}_0\left(2.78-\frac{3}{2}\ln
 \frac{M_\Ups}{\mu}\right) -14.1\right)\right]\,.\eeq
 } 
 Here, $\tilde{\beta}_0=11-\frac{2}{3}n_f$ (in
 PDG normalization) with $n_f=5$ as \msbar scheme is used.\par

   Meanwhile, in the RG-invariant expression $R(\mu^2)\,,$
 scale $\mu$ can appear {\it only} in the argument of
 QCD coupling $\as(\mu^2)\,.$ To return eq.(\ref{Mack})
 to the RG-invariant form, one could set $\mu=M_{\Ups}\,$
  {\small \beq\label{R-Yps}
 R_\Ups=
 \frac{10(\pi^2-9)\as^3(s_\Ups)}{9\,\pi\alpha^2(M_\Ups)}
 \left(1+\frac{\as(s_\Ups)}{\pi}\, 7.2\right)=
 5360[\as^3(s_\Ups)+2.30\as^4(s_\Ups)]\,; \quad
 s_\Ups=M^2_\Ups\,.\eeq}
  Then, the issue of scale should be readdressed to the
 choice of $s_\Ups\,$ in eq.(\ref{R-Yps}).

 It was argued~\cite{kobelthesis} that a proper value for
 scale $\mu$ could be $M_\Ups/3\,,$ due to the 3-gluonic
 mode of the $\Ups$ decay. Below, we consider the range
 close to $M_\Ups\,,\sqrt{s_{\Ups}}=7-9$ \GeV, and discuss
 the scale uncertainty in the final error estimate.
 The QED coupling $\,\alpha(M_\Ups)$ is fixed on the
 $\Ups$ mass.\par

\subsection{Reevaluation of $\Lambda_{QCD}$ from $\Ups$
 Decay} 
 \ \ \ \underline{Calculations.} The first attempt to
 re-evaluate $\Lms$ extracted from $\Ups$ decay by proper
 taking into account analytic continuation effects was
 made\footnote{One should remark here that some other
 kind of summation, namely of $1/v-$type terms, for the
 improvement of $\Lms$ extraction from $\Ups$ production
 cross-sections was devised in \cite{Pivov98a}. Unlike
 the present case, where we deal with $R_Y$ for $\Ups$
 decay, the authors of \cite{Pivov98a} studied sum rules
 for ratio $R(s)=\frac{\sigma(e^+e^-\to b\bar{b})}
 {\sigma(e^+e^-\to \mu^+\mu^-)}$. Analytic continuation
 effects were not taken into account.} in
 \cite{KrasPiv82}. \par
  Analysis analogous to \cite{KrasPiv82} is performed
 here, with the APT expansion
 \beq\label{UpsAPT}  
 R_\Ups^{theor}=5360\left[\agoth_3(L)+2.30\,
 \agoth_4(L)\right]\eeq
 instead of formula~(\ref{R-Yps}) which, in turn,
 within our Model, looks like
\begin{equation}\label{UpsMod}
 R_\Ups^{Mod}=5360\left[\agoth_3^{Mod}(L)+
            2.30\,\agoth_4^{Mod}(L)\right]. \eeq
 By these formulae we extract $\Lms^{(5)}$ and \as
 values from fresh CLEO III\cite{Adams:2004xa} data
\begin{equation}\label{cleo}
  R_\Ups^{CLEO}=37.3\pm 0.75\end{equation}
 A few words on the fourth APT function $\agoth_4(L)\,.$
 In the r.h.sides of eq.(\ref{UpsAPT}) we use
 unpublished yet numerical results \cite{MagrPriv} and
 in (\ref{UpsMod})-- the Model expressions, eqs.
 (\ref{2-3}), (\ref{4m}), (\ref{model}) with
 logarithmic argument (\ref{model2}):
 $\,\agoth_{3,4}^{Mod}(L)=\agoth_{3,4}^{(1)}(L_2)\,;
 \quad \,L_2= L+b\ln\sqrt{L^2+2\pi^2} \,.$\smallskip

   As it follows from a more detailed
 analysis\footnote{Based upon calculation\cite{MagrPriv}
 of exact APT formula.}, at $\sqrt{s}\sim 7-9\,\GeV\,$
 relative error of $\agoth_3^{Mod}(L)\,$ is about 6\,\%.
 At the same time, in this interval, the ratio
 $|\agoth_4^{Mod}/\agoth_3^{Mod}|\sim 0.16\,.$ This means
 that the error due to the 2nd term in the r.h.s. of
 eq.(\ref{UpsMod}), is about 3\,\% and the total Model
 error for $R_\Ups^{Mod}\,$ about 9\,\%. In turn, this
 gives 3 \% Model error for \as that is equal to
 $\triangle \as(M_\Ups)= \pm 0.005\,.$

  Now, numerical analysis of the "Exact" case yields
 $\Lambda_5= 210\pm 5\,\MeV\,,$ that is $\as(M_Z)=
 0.118\pm 0.0005\,.$ Further on, we shall neglect by
 this small error. For the "Model" case with
 $\Lambda_5= 235,\MeV\,,$ that is $\as(M_Z)=0.120\,,$
 one has the Model error $\triangle\as(M_Z)=\pm 0.002\,.$

 In line 1 of Part I, ``PDG, $1S$", standard PT results on
 $\Ups(1S)$ decay are given. We present them not exactly
 as they were published in~\cite{pdg04} but recalculated
 along with modern experimental data. Line 2, marked
 ``PDG, Fit" gives the published world average value
 described by the curve on Fig. 9.2 in \cite{pdg04},
 within the error bars of {\it all} the processes. Column
 ``$\alpha(M_\Ups)$" means ``$\as$,  calculated at the
 mass of $\Ups$, according to eq.(9.5) of \cite{pdg04}".\par
  Line 1 of Part II, ``Exact APT", presents results of
 $\Ups$1S decay calculated by exact numeric tables for
 $\agoth_3\,$ and by \cite{MagrPriv} for $\agoth_4\,$ APT
 function.
 Line 2, [Mod], presents values obtained from $\Ups(1S)\,$
 decay data by means of the Model eqs.(12),(14). Here, model
 errors combine Model errors of both the terms in the r.h.s.
 of eq.(18).
 Line 3, ``Crude APT" gives an earlier result\cite{Sh01epjc}
 with approximate APT estimate used to correct (Corr) the
 Bethke-2000 value $\as(M_\Ups)=0.170\,$ extracted there
 from all the $\Ups$ decays data. \par

 Everywhere in Part II, we translate our results from
 APT expressions, by use of related $\Lambda_{5}\,$
 values into standard {\it non-analytic} coupling \as
 at $M_\Ups$ and $M_Z$ for ease of comparison with Part I
 and other standard sources.\par 

 The importance of taking into account the analytic
 effects for proper $\as$ extraction in the low-energy
 range is clearly seen from the Table 3. It is also evident
 that the Model error is small enough even in the
 non-favorable case of quarkonium decay, in particular when
 compared with the scale error.

   \underline{The scale uncertainty.} Besides ``direct" data
 error that in our case is 2\% for $R_\Ups\,,$ there is a
 ``hidden" uncertainty in theoretical equations (\ref{UpsAPT})
 and (\ref{UpsMod}) related with a choice of argument
 $L=\ln(s_\Ups/\Lambda^2)\,.$ In given figures, we use
 $s_\Ups=M^2_\Ups\,.$ To discuss effect of the $s_\Ups\,$
 variation, return to issue mentioned above at the end of
 Sect.4.1. .\par
  This reference scale issue is actual for QCD analysis of
 all low energy data. The RG non-invariant term
 $\ln(M_\Ups/\mu)\,$ in the r.h.s. of eq.(\ref{Mack}) just
 represents an attempt to take into account this effect. The
 scale effect enlarges quickly with the uncertainty rise.
 E.g., for its value 1 GeV, i.e., for 8--9 \GeV interval,
 we have  $205\leq\Lambda_5 \leq 240\,\MeV\,,$ while to the
 7--9 \GeV case --- $180 \leq \Lambda_5 \leq 240\,\MeV\,.$
 In the Table 3, where the APT results are compared with
 ones of standard PT, for the scale error $[...]_{sc}\,$
 we conditionally give figures related
 to\footnote{To the case $\sqrt{s_\Ups}=M_\Ups/2=M_b\,$
 there corresponds $\alpha_s(M_Z)\sim 0.114\,.$ and
 to  $\sqrt{s_\Ups}= M_\Ups/3\,$\cite{kobelthesis}
 -- $\alpha_s(M_Z)\sim 0.110\,.$} the 2 \GeV interval.\par
  Generally, the scale issue is an intrinsic problem of
 renormalization group application to observables.
  On the one hand, the common, {\it``vulgaris"}, version of
 RG algorithm corresponds to the UV (massless) case with
 simplified definition of effective (running) coupling
 $\albar(q^2/\mu^2)\,.$ Here, it is tacitly assumed that
 vertex function $\Gamma(q^2_1,q^2_2,q^2_3)\,$ entering
 into the \albar definition is taken with equal arguments.
 On the other hand, the condition $q^2 \gg m^2\,$ is used.
 Both the assumptions are not valid for the low energy QCD
 case. On these items, the reader could be addressed to
 mass-dependent RG formalism \cite{dv92-3,dv92-4,dv95dok}.

\section{Conclusion}
\begin{enumerate}
\item
 Our theoretical result is model explicit expressions
 (\ref{model}) for the analyticized 3-loop couplings
 (\ref{AE11}),(\ref{AM11}) and their effective powers
 (\ref{2-3}),(\ref{4m}) in both the Euclidean and Minkowskian
 regions. These are just one-loop analytic expressions of
 Analytic Perturbation Theory with modified logarithmic
 arguments (\ref{model}). The accuracy of the Model is
 estimated to be sufficient for practical purposes in the
 region 1--100\,GeV, which hosts many important processes. \par
\item
 To illustrate the APT and Model application, we consider the
 $\as$ value extraction from data for $\Ups$(1S) non-radiative
 decay measured with 2\% error. Fixing the scale at the $\Ups$
 mass, we got (for the $\as^{\MSbar}\,$ coupling) by exact
 APT numerical calculation and by our APT Model
\begin{equation}\label{APTres}
\alpha_s^{\rm APT exact}(M_Z) =0.118(1)_{exp}\,,\quad
 \alpha_s^{\rm APT Model}(M_Z) =0.120(2)_{Mod}\,.\eeq
 The comparison with result of $\Ups$(1S) usual analysis,
 $\alpha_s(M_Z)=0.112(2)\,,$ and with the world average,
 $\as(M_Z)=0.1185(20)\,,$  confirms the validity of the
 devised Model.
 \item
 It was established that the scale uncertainty essentaially
 reduces the value of theoretical analysis of $\Ups$ decay.
 In the considered case, the scale error dominates over
 experimental and Model ones. This issue is worth urgent
 further examining.
 \end{enumerate}

\section*{Acknowledgements}
 The authors are grateful to A.Bakulev, A.Kataev, S.Mikhailov,
 A.Pivovarov for important advices, to A.Efremov, S.Gerasimov,
 F.Jegerlehner, N.Krasnikov, V.Rubakov, I.Solovtsov,
 N.Stefanis, O.Tarasov, and O.Teryaev for stimulating
 discussions, as well as to R.Pasechnik for help in numerical
 calculation. Special thanks are addressed to B.Magradze for
 performing exact computation and providing us with tables
 of the forth APT functions. \par

 We are glad to thank also R.M. Barnett, S.Eidelman,
 I.Hinchliffe, and Weiming Yao for their attention to our
 remarks concerning $\Ups$ data in PDG review and for taking
 them into account in the 2005 online version.
 One of us (A.Z.) thanks M.Kobel for kindly sending his
  Thesis and clearing some points.
  The work has been supported in part by RFBR grant No.
 05-01-00992, Scient.School grant 2339.2003.2 and by JINR
 Heisenberg --- Landau project. One of the authors (A.Z.)
 was supported in part by DAAD Forschungsstipendium.\par


\newpage
 \begin{table}[h]\label{prec} 
 \caption{\bf Maximal errors of the Model with $a=2$ (old)}%
\begin{center}
\begin{tabular}{|c||c|c|c|c|c|c|}\hline
 $\hspace{12mm}n_f$& $Er\agoth_1$& $Er\agoth_2$& $Er\agoth_3$&
 $Er\acal_1\,$&$Er\acal_2\,$& $Er\acal_3\,$\\ \hline
 3 {\small(0.5 - 1.0\.GeV)}&3\,\%&5\,& 15 \,&3\,&10\,&20\,\\ \hline
 3 {\small(1 - 1.5\.GeV)}&3\,\%&4\,& 8 \,& 2\,&7\,&10\cr \hline
 4 &1\,\%&3\, &8 & 2 & 10\,& 20\,\\ \hline
 5&1\,\%&3&6\,&0.5\,&4\,&15\, \\ \hline
\end{tabular}
\end{center}
 \end{table} 

  \begin{table}[h]\label{prec} 
 \caption{\bf Maximal errors of the Model with $a=2$ (corrected)}%
\begin{center}
\begin{tabular}{|c||c|c|c|c|c|c|}\hline
 $\hspace{12mm}n_f$& $Er\agoth_1$& $Er\agoth_2$& $Er\agoth_3$&
 $Er\acal_1\,$&$Er\acal_2\,$& $Er\acal_3\,$\\ \hline
 3 {\small(0.5 - 1.0\.GeV)}&3\,\%&17\,& 39 \,&5\,&16\,&8\,\\ \hline
 3 {\small(1 - 1.5\.GeV)}&4\,\%&5\,& 9 \,& 3\,&7\,&7\cr \hline
 4 &0.5\,\%&3\, &9 & 1 & 10\,& 21\,\\ \hline
 5 &1\,\%&3&6\,&0.5\,&4\,&14\, \\ \hline
\end{tabular}
\end{center}
 {\footnotesize Notation: chan. -- \ t = Euclidean or $s=$
 Minkowskian channel; scale $q=\sqrt{s}$ or
  $q=\sqrt{Q^2}\,;$ \\ $\Delta \alpha/\alpha$ -- relative
 error of $\alpha$ at a given scale; DIS (Bjork.) --
 \ Bjorken scaling violation in DIS, \\ GLS  --- \  Gross --
 Llewellyn-Smith sum rule, and $^*$ marks combined theor.
 and exper. errors.}
\end{table}

\newpage
\begin{table}[h]  
 \caption{\bf Results of various $\as$ extraction
   from Upsilon decays}
\medskip
\begin{center}
\begin{tabular}{|l|c|c|c|} \hline
\multicolumn{4}{|c|}{Part I. Non-APT treatment}\cr \hline
\multicolumn{1}{|c|}{Source}&$\alpha(M_\Ups)$ &$\alpha(M_Z)$&
$\Lms^{n_f=5}$\cr \hline PDG, $\Ups,1S$
&0.170(4)&0.112(2)&146$^{+18}_{-17}$  \cr \hline PDG, global
Fit&0.182(5)& 0.1185(20)&217$^{+25}_{-23}$\cr \hline\hline
\multicolumn{4}{|c|}{Part II. APT treatment}\cr\hline
 Exact APT,\,1S&0.1805$(12)_{exp}$&0.1179$(5)_{exp}$&210(5)\cr\hline
\bf [Mod],\,$1S$ & \bf 0.185$(5)_{\rm M}$&
 0.120$(2)_{\rm M}$& 235$(25)$\cr\hline
\small\cite{Sh01epjc} Crude APT&0.183 & 0.119 &222\cr\hline
 \small Exact$\pm${\sf [scale]},\,$1S$&0.180$[7]_{sc}$ &
 0.118$[3]_{sc}$&210$[30]$\cr\hline
\end{tabular}\end{center}
\end{table}

\end{document}